\documentclass[conference]{IEEEtran}
\IEEEoverridecommandlockouts
\usepackage{cite}
\usepackage{amsmath,amssymb,amsfonts}
\usepackage{algorithmic}
\usepackage{graphicx}
\usepackage{textcomp}
\usepackage{xcolor}
\usepackage{listings}
\usepackage{booktabs}
\usepackage{booktabs} 
\usepackage{array}
\usepackage{lipsum} 
\usepackage[utf8]{inputenc}
\usepackage{array}
\usepackage{verbatim}
\usepackage{url}
\usepackage{float}
\usepackage{capt-of}
\newcolumntype{M}[1]{>{\arraybackslash}m{#1}}
\newcolumntype{L}[1]{>{\arraybackslash\raggedright}m{#1}}

\usepackage{enumitem}
\newlist{tblitemize}{itemize}{1}
\setlist[tblitemize,1]{label=\labelitemi, nosep, topsep=0pt, partopsep=0pt, leftmargin=10pt, labelindent=0pt, itemindent=0pt}

\newcolumntype{P}[1]{>{\raggedright\arraybackslash}p{#1}}

\def\BibTeX{{\rm B\kern-.05em{\sc i\kern-.025em b}\kern-.08em
    T\kern-.1667em\lower.7ex\hbox{E}\kern-.125emX}}
\begin{document}

\title{Spatial Reasoner: A 3D Inference Pipeline \\ for XR Applications}

\author{
\IEEEauthorblockN{Steven Häsler\textsuperscript{1} \\
hasv@zhaw.ch \\
} \\
\IEEEauthorblockA{\textsuperscript{1}\textit{ZHAW Zurich University of Applied Sciences} \\
\textit{School of Engineering}\\
Winterthur, Switzerland
}
~\\
\and
\IEEEauthorblockN{Philipp Ackermann\textsuperscript{1,2} \\
acke@zhaw.ch, philipp@metason.net
} \\
\IEEEauthorblockA{\textsuperscript{2}\textit{Metason} \\
Zurich, Switzerland
}
}

\maketitle

\begin{abstract}
Modern extended reality (XR) systems provide rich analysis of image data and fusion of sensor input and demand AR/VR applications that can reason about 3D scenes in a semantic manner. We present a spatial reasoning framework that bridges geometric facts with symbolic predicates and relations to handle key tasks such as determining how 3D objects are arranged among each other ('on', 'behind', 'near', etc.). Its foundation relies on oriented 3D bounding box representations, enhanced by a comprehensive set of spatial predicates, ranging from topology and connectivity to directionality and orientation, expressed in a formalism related to natural language. The derived predicates form a spatial knowledge graph and, in combination with a pipeline-based inference model, enable spatial queries and dynamic rule evaluation. Implementations for client- and server-side processing demonstrate the framework’s capability to efficiently translate geometric data into actionable knowledge, ensuring scalable and technology-independent spatial reasoning in complex 3D environments. The Spatial Reasoner framework is fostering the creation of spatial ontologies, and seamlessly integrates with and therefore enriches machine learning, natural language processing, and rule systems in XR applications. 
\end{abstract}


\section{Introduction}
Spatial computing, which includes various immersive technologies such as extended reality (XR), augmented reality (AR), virtual reality (VR) and mixed reality (MR), merges the digital and physical worlds.
Sensor fusion of imaging cameras, depth sensors (e.g. LiDAR) and inertial measurement units (IMUs) provides 3D data that is further processed using computer vision and machine learning methods to enable seamless interaction between the digital and physical worlds. 

Extended reality applications need to understand the physical world to overlay digital objects that appear three-dimensional and blend with the real environment. The application logic of such XR applications needs to deal with real and virtual objects. The dynamic nature of captured, real and of physically simulated or user-manipulated virtual objects in time and space makes programming quite challenging.   

\section{Motivation}

The goals of the presented Spatial Reasoner are manifold: a) Overcoming the limitations of hard-coded 3D application logic through a general representation of 3D spatial knowledge, b) Providing a comprehensive and systematically constructed collection of spatial predicates and corresponding spatial relations for 3D sceneries, c) Using a natural language-like formalism (instead of mathematical abstract declarations) to simplify the formulation and readability of spatial knowledge and to enable easy integration with NLP and LLMs, d) Supporting a technology-independent and cross-platform approach.  

Spatial fuzziness affects information retrieval in space. Object detection in state-of-the-art computer vision, machine learning, and Augmented Reality toolkits results in detected objects that constantly vary their locations and do track and improve over time their orientations and boundaries in space. The object description is to a certain level fuzzy and imprecise, yet some non-trivial conclusion can anyhow be deduced. The geometric confidence typically improves over time. Additionally, by taking spatial domain knowledge into account, semantic interpretation and therefore overall confidence of scene understanding can be increased. The goal of the Spatial Reasoner library is to improve object recognition using domain knowledge and taking into account spatial-semantic and three-dimensional topological conditions.

\section{Related Work}

Spatial queries and spatial reasoning are critical components across a wide range of disciplines: from geographic information systems (GIS) and building information modeling (BIM) to augmented reality. Early foundational work in spatial relationships established three core categories: topological relations (\textit{meet}, \textit{overlap}, \textit{disjoint}) formalized through models like the 9-intersection framework \cite{Zuliansyah_2008_3DTopological}), metric-based (e.g., distance thresholds \cite{Taniar_2013_SpatialQueries}), and directional relations (e.g., cardinal directions \cite{Wang_2024_3DCardinalDirectionInverse, Carniel_2023_DefiningQueries}). Such definitions enable queries like ``find all roads intersecting region X'' or ``retrieve polygons within 3\,km of a point of interest.''

When moving to 3D, the complexity increases significantly \cite{Santos02012025}. Researchers have developed refined cardinal direction models, such as the 3DR27 formalism \cite{Wang_2024_3DCardinalDirectionInverse} and 3D topological expansions of standard 9-intersection approaches. These expansions allow queries like ``find all building floors directly above the parking garage'' or ``are there support columns behind the east wall?''  Although they provide greater richness in modeling, they also raise new challenges for spatial indexing, representation, and reasoning, particularly for consistency checks \cite{Wang_2024_BlockAlgebraConsistency}.

Another long-standing thread is the inclusion of spatial relationships in relational database systems and geographic information systems (GIS) \cite{forrest2023spatial}, along with the relevant index structures \cite{Carniel_2023_DefiningQueries}. This approach enables user-friendly retrieval in large geospatial databases or big 3D data from BIM. For example, PostGIS and other systems supply topological and distance-based operators. More complex 3D directional queries typically require specialized engines or external libraries.

Modeling spatial facts in knowledge graphs (or as semantic web) is a well-known technology for spatial reasoning  \cite{Zhang_Lu_Du_Chen_Cao_2021, MANTLE2024125115}.
Knowledge graphs represent entities and relations as computable networks. For spatial reasoning, entities represent 2D or 3D objects, and relations represent how the objects relate to each other in space.

Recently, bridging the gap between knowledge representation and geometry is addressed in systems like Clingo2DSR, which merges Answer Set Programming with advanced geometry computations \cite{Clingo2DSR}. 
Similarly, in laparoscopic or surgical settings, scene analysis may require dynamic relationships such as ``landmark is obstructed behind tissue'' and ``tool is oriented over an organ region'' \cite{GraphBasedSpatReason}. Data-driven models have been proposed to discover 3D relationships from real-world distributions (e.g., \emph{on}, \emph{next-to}, \emph{aligned}) \cite{Li_2025_Generalize3D}.

In addition, graph-based methods such as RGRN capture semantic and spatial relationships through relation graphs to enhance object detection\cite{zhao2023rgrn}, while others leverage proximity cues from scene understanding to construct a spatial knowledge base \cite{xu2024hierarchical}. Open-vocabulary object placement \cite{Sharma_2024_OCTOPLUS} is also increasingly being investigated, leveraging cardinal directions or adjacency constraints so that AR or VR systems can place content in contextually relevant ways. Moreover, recent work in vision language models, such as SpatialVLM \cite{chen2024spatialvlm}, demonstrates that incorporating large-scale synthetic spatial reasoning data can endow VLMs with quantitative spatial reasoning capabilities by enabling metric distance estimation and chain-of-thought reasoning for applications in robotics and AR.

Overall, from 2D topological queries to 3D cardinal direction expansions to open-vocabulary AR scenarios, the literature converges on the idea that robust, multi-faceted definitions of spatial relationships yield richer query types and better solutions. Yet it also indicates that additional synergy is needed between geometry-based formalisms (e.g., the 3DR27 model) and reasoning frameworks (e.g., ASP-based \cite{Clingo2DSR}), to ensure tractable consistency checking, symbolic spatial rules, and dynamic or uncertain scenario handling. Computation of spatial reasoning has recently undergone improvements and it can be found in multiple application domains:

\noindent\textbf{(1) Geographic Information Systems.} 
Geographic Information Systems (GIS) provide spatial analysis based on geometric and cartographic characteristics. Spatial queries in databases for geostatistics are supported by Spatial SQL \cite{forrest2023spatial}. GeoSPARQL 
\cite{nicholas_j_car_ogc_2023} defines an extension to the SPARQL query language for processing of geospatial data in RDF tripe stores.

\noindent\textbf{(2) Architectural Applications and BIM.} 
In code compliance checking, complicated rules such as ``no direct line of sight between an access route and a bathroom fixture'' can be defined as relationships in 3D \cite{Clingo2DSR}. Similar constraints like ``no overlap in bounding volume'' or ``pillar is strictly above floor slab'' require cardinal direction or topological checks across large building models. The 3DR27 model \cite{Wang_2024_3DCardinalDirectionInverse}, with formal definitions for 3D direction relations, can facilitate straightforward query statements (``above,'' ``beside,'' ``behind''), bridging domain-level constraints (like safety rules) with geometric computations.

\noindent\textbf{(3) Robotics Scene Understanding.}
By using scene understanding \cite{slam22197265}, a robot might reason in a cluttered environment: ``object on top of the table, aligned with or behind a shelf.'' Classic bounding box or 2D topological logic can fail to capture real 3D interactions or minor orientation differences \cite{Li_2025_Generalize3D}. By using advanced cardinal direction or block algebra \cite{Wang_2024_BlockAlgebraConsistency}, the robot can filter out impossible states and handle large-scale consistency checks. Meanwhile, symbolic reasoning frameworks unify domain constraints and geometry computations, enabling the robot to place items stably or interpret partial occlusions \cite{LIU2023104294}.

\noindent\textbf{(4) XR/AR/VR and Mixed Reality.}
Determining the dynamic situation of detected, virtual, simulated and manipulated 3D objects in XR applications has been addressed by integrating deep learning, semantic web and knowledge graphs \cite{LAMPROPOULOS202032}, \cite{Muff2022AFF}.
Systems like OCTO+ \cite{Sharma_2024_OCTOPLUS} require placing virtual objects in physically plausible or semantically coherent positions. For example, a painting is placed on a \emph{vertical} plane, near or above a piece of furniture. 3D direction constraints help define valid surfaces or adjacency conditions, improving user experiences for open-vocabulary content (``place a cupcake on the nearest table'').

In short, deeper integration of these advanced 3D relationships foster automated, user-friendly queries. By combining efficient indexing or incremental search with cardinal direction constraints, domain experts can formulate queries in terms of \emph{north-of, behind, above} or \emph{inside} relationships, rather than manually coding geometric transformations. This impetus is only growing, as more systems adopt 3D data, from city-scale digital twins to increasingly sophisticated robotics tasks and interactive XR applications.

\section{Spatial Reasoner}
\label{sec:spatial-reasoning}
Spatial reasoning is the ability to understand and interpret the positions and relationships of objects in three-dimensional space. In many domains, such as augmented reality (AR), robotics, architecture, and gaming, applications must not only detect and track objects but also infer how those objects are arranged and interact. The Spatial Reasoner\footnote{\url{https://github.com/metason/SpatialReasoner}} presented in this work enables such high-level inference by combining  bounding-box geometry, deducing spatial attributes and relations, and rule-based processing.

\subsection{Spatial Objects}
\label{subsec:spatial-objects}

The spatial Reasoner operates on a \emph{fact base} consisting of one or more \emph{spatial objects} that can be related to each other. These objects can either represent objects:
\begin{itemize}
    \item \textbf{Detected in the real world}, via 3D sensors such as camera-based computer vision, machine learning (ML) engines, and point-cloud processing libraries, or 
    \item \textbf{Generated virtually}, for instance within a 3D scene graph or a physical simulation environment.
\end{itemize}

The Spatial Reasoner focuses on oriented bounding boxes rather than high-fidelity meshes. The spatial objects are loaded into the reasoner which will analyze or augment the objects and their relations.
Each \lstinline|spatial object| represents a real or virtual entity in space by means of an \emph{oriented bounding box} (bbox). The bounding box is axis-aligned to the horizontal ground plane and rotated around the centered up vector. Concretely:
\begin{itemize}
    \item The \textit{rotation} in the horizontal plane is provided as an \lstinline|angle| in \emph{radians}, measured counterclockwise when viewed from above. A value of \emph{yaw} is calculated in degrees for user-friendly displays.
    \item The \textit{position} \((x, y, z)\) denotes the center of the bounding box’s \textit{base area}—also called the “footprint.”
    \item The \textit{extent} of the box is defined by \lstinline|width| (\(w\)), \lstinline|height| (\(h\)), and \lstinline|depth| (\(d\)).
\end{itemize}

The position and rotation of the object around its vertical axis may differ depending on the orientation of the underlying left- or right-handed coordinate system. The concrete implementation of a Spatial Reasoner library abstracts such differences but \emph{requires} the user to adopt a consistent reference frame for accurate inference.

\subsection{Attributes of Spatial Objects}
\label{subsec:key-attributes}

Each \lstinline|spatial object| must specify a unique \textit{id}, which needs to be kept constant if the object is updated over time. Table~\ref{tab:spatial-attributes} summarizes the key attributes expected by the Spatial Reasoner:

\begin{table}[htbp]
\caption{Core Attributes of a SpatialObject}
\label{tab:spatial-attributes}
\centering
\begin{tabular}{l p{5.5cm}}
\hline
\textbf{Attribute} & \textbf{Description} \\
\hline
\textit{id} & Unique identifier for tracking the same object across updates \\
\textit{x, y, z} & Position of the object’s base center in 3D space \\
\textit{w, h, d} & Width, height, and depth (box extents) \\
\textit{yaw} & Rotation around the up axis given in radians \\
\hline
\end{tabular}
\end{table}

These attributes form the minimal specification required for spatial reasoning. Additional fields (\textit{label}, \textit{confidence}, \textit{velocity}, etc.) can further refine the properties of each object, supporting more specialized inferences in scenarios such as object classification, motion analysis, or user interaction tracking.

From declared object attributes, the Spatial Reasoner automatically deduces metric attributes (\textit{footprint area}, \textit{volume}, \textit{perimeter}, etc.) as well as boolean attributes (\textit{equilateral}, \textit{thin}, \textit{moving}, etc.).

\subsection{Spatial Reference Systems}
\label{subsec:ref-sys}

The interpretation of some predicates of spatial relations depends on the frame of reference. E.g., predicates such as left, right, in front, and at back have different meanings in different reference systems. Furthermore, the semantics of spatial predicates in English is sometimes vague and it is hardly possible to distinguish between terms and their synonyms (e.g., over, above, ontop). Therefore, the meaning of all spatial predicates used in the Spatial Reasoner library is clearly defined (see appendix). Although the ordinary meaning of the terms has been taken into consideration, the specification in the Spatial Reasoner library may not correspond to everyday usage in spoken English.

The interpretation of spatial predicates and their corresponding relations are only valid in certain reference systems:

\begin{itemize}

    \item \textbf{World Coordinate System (WCS)}: spatial relations are encoded relative to a global reference point and the orientation of its coordinate system.
    \item \textbf{Object Coordinate System (OCS)}: spatial relations are encoded relative to the local position and orientation of an object
    \item \textbf{Egocentric Coordinate System (ECS)}: spatial relations are encoded relative to the position and view direction of an observer
    \item \textbf{Geodetic Coordinate System (GCS)}: spatial relations are encoded relative to earth's projected latitude (north/south) and longitude (east/west)
 \end{itemize}
   
\subsection{Spatial Sectors}
\label{subsec:spatial-sectors}

In order to reason about the \textit{relative positioning} of objects, the Spatial Reasoner partitions the space around each reference object into \emph{sectors}. Conceptually, the bounding box is extended along the three orthogonal axes—left-right, ahead-behind, and over-under—to form a \(3 \times 3 \times 3\) grid of \emph{27  sectors}.

\begin{itemize}
    \item \textbf{Zero Divergency} (\texttt{i}) for inner/inside sector
    \item \textbf{Single Divergency} (\texttt{l, r, a, b, o, u}) denotes a single principal direction offset (e.g., an object in sector \texttt{l} is within the left sector of the reference bbox).
    \item \textbf{Double Divergency} (\texttt{al, ar, bl, br, lo, lu, ro, ru, ao, au, bo, bu}) captures a combination of two directions (e.g., \texttt{ar} means “ahead-right,” \texttt{lu} means “left-under”).
    \item \textbf{Triple Divergency} (\texttt{alo, aru, blu, bru}, etc.) represents more nuanced positions such as “ahead-left-over” or “behind-right-under.”
\end{itemize}

\begin{figure}[htbp]
\centerline{\includegraphics[width=0.5\textwidth]{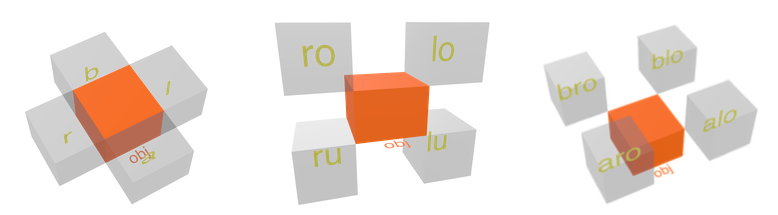}}
\caption{Single, double, and triple divergency of sectors.}
\label{divergency}
\end{figure}

\begin{figure}[htbp]
\centerline{\includegraphics[width=0.5\textwidth]{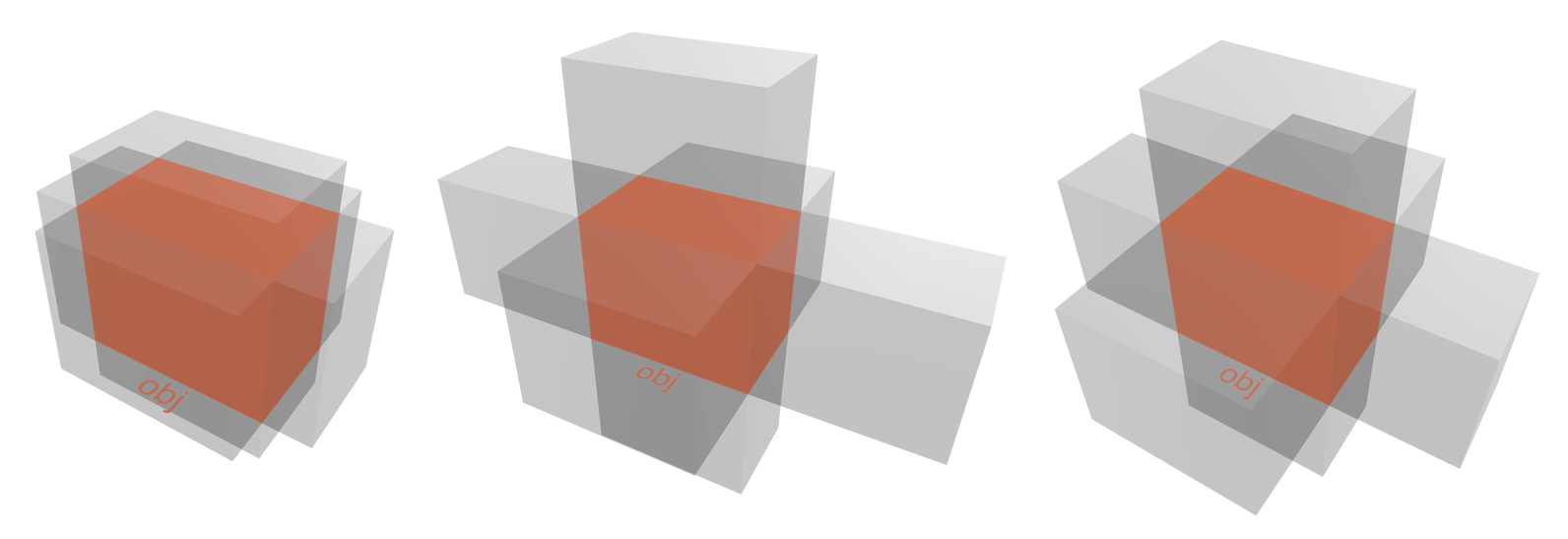}}
\caption{Samples of adjustable sector sizes.}
\label{sectors}
\end{figure}

By dividing space into these discrete regions, the Spatial Reasoner systematically assigns objects to appropriate sectors and infers adjacency relationships. Different \texttt{sector schema} options (e.g., \texttt{fixed}, \texttt{dimension}, \texttt{nearby}) and a \emph{sector factor} can customize the size of each sector region (Fig.~\ref{sectors}), ensuring it fits the scale of the environment (e.g., in a large warehouse vs.\ in a small indoor room).

\subsection{Spatial Predicates and Spatial Relations}
\label{subsec:spatial-predicates}

Beyond bounding-box geometry, the Spatial Reasoner supports a rich set of \emph{qualitative predicates} that formalize on a symbolic level how objects relate to each other. Spatial relationships are represented as \textbf{\emph{subject - predicate - object}} patterns. The following spatial relation categories are supported:
\begin{itemize}
    \item \textbf{Proximity}: \emph{near}, \emph{far}
    \item \textbf{Directionality}: \emph{left}, \emph{right}, \emph{ahead}, \emph{behind}, \emph{above}, \emph{below}
    \item \textbf{Adjacency}: \emph{leftside}, \emph{frontside}, \emph{beside}, \emph{ontop}, ...
    \item \textbf{Orientation}: \emph{aligned}, \emph{orthogonal}, \emph{opposite}, ...
    \item \textbf{Connectivity}: \emph{on}, \emph{in}, \emph{by}, \emph{at}
    \item \textbf{Sectoriality}: \emph{i, o, u, al, alo, ...} (referring to bbox sectors)
    \item \textbf{Assembly}: \emph{disjoint}, \emph{inside}, \emph{touching}, \emph{meeting}, ...
    \item \textbf{Visibility}: \emph{seenleft}, \emph{infront}, ... (relative to an observer)
    \item \textbf{Comparability}: \emph{bigger}, \emph{smaller}, \emph{longer}, \emph{shorter}, ...
    \item \textbf{Similarity}: \emph{samewidth}, \emph{samevolume}, \emph{congruent}, ...
    \item \textbf{Geography}: \emph{north}, \emph{south}, \emph{northwest}, ...
\end{itemize}

Each predicate has an underlying geometric or semantic definition. For example, \emph{near} and \emph{far} rely on distance thresholds defined by the \texttt{nearby} schema, whereas \emph{on}, \emph{in}, and \emph{by} infer direct contact or containment taking a \texttt{max gap} into consideration. Meanwhile, \emph{seenleft} and \emph{infront} consider the point of view of an observer to relate two objects. Details on spatial predicates and relations can be found in the appendix.

By combining multiple predicates, the system can draw higher-level conclusions, such as:
\begin{quote}
\textit{``If an object is near and in front, it may be physically reachable by the user.''}
\end{quote}
or
\begin{quote}
\textit{``If two objects are touching and one is above the other, we can classify this as a stack or a pile.''}
\end{quote}

Such qualitative interpretations on a symbolic level are invaluable in tasks like navigation, scene analysis, object classification, or spatial aggregation where strictly numeric bounding-box comparisons can be cumbersome or insufficiently descriptive.

\section{Spatial Inference Pipeline}
\label{sec:inference-pipeline}

The spatial inference pipeline converts the processing of object attributes and predicates into a linear, rule-based workflow. At each stage, it relates, refines, or transforms a collection of spatial objects stored in the current context. 

\subsection{Pipeline Specification}

Users specify pipelines as a \emph{textual} sequence of operations, separated by the pipe character '\texttt{|}'. Each operation follows an \textbf{\emph{input--process--output}} pattern:
\begin{itemize}
    \item \textbf{Input}: A list of object references 
    \item \textbf{Process}: A specific operation (e.g., \texttt{filter}, \texttt{pick}, \texttt{sort}, etc.)
    \item \textbf{Output}: A modified list or the same list with different ordering or changed object attributes
\end{itemize}

Formally:
\[
    \text{pipeline} : \text{operation}_1 \,|\, \text{operation}_2 \,|\, \dots \,|\, \text{operation}_n
\]
The pipeline begins with all objects loaded in the fact base and ends with a potentially changed subset or superset, depending on the operations executed.

\subsection{Inference Pipeline Operations}

The core operations include:
\begin{itemize}
    \item \texttt{\textbf{adjust}}: Sets contextual parameters (e.g., \emph{nearby thresholds}, \emph{sector dimensions}, \emph{max deviation}, \emph{long/thin ratios}).
    \item \texttt{\textbf{deduce}}: Defines which relation categories (e.g., \emph{topology}, \emph{connectivity}, \emph{visibility}) will be computed.
    \item \texttt{\textbf{filter}}: Removes objects failing a specified attribute condition (e.g., \texttt{width > 0.5}).
    \item \texttt{\textbf{isa}}: Choose objects that belong to a type in a class hierarchy (a taxonomy of spatial objects).
    \item \texttt{\textbf{pick}}: Collects objects based on certain spatial relations (e.g., \emph{left}, \emph{near}, \emph{above}) that exist with input objects.
    \item \texttt{\textbf{select}}: Similar to SQL, selects objects that have a specified relation to other objects.
    \item \texttt{\textbf{sort}}: Orders objects by numeric attributes (\texttt{volume}, \texttt{width},...) or relation-based metrics (e.g., \texttt{distance}).
    \item \texttt{\textbf{slice}}: Restricts the current set to a subsection (e.g., the first object or an index range).
    \item \texttt{\textbf{calc}}: Calculates global variables (e.g., \texttt{count}, \texttt{median}, \texttt{max}) for use in subsequent steps.
    \item \texttt{\textbf{map}}: Updates or creates local object attributes (e.g., \texttt{shape = 'cylindrical'}).
    \item \texttt{\textbf{produce}}: Generates new spatial objects driven by a relation, such as placing a corner object where two walls meet.
    \item \texttt{\textbf{reload}}: Reloads all objects of the fact base as input to the next pipeline operation, including newly produced ones.
    \item \texttt{\textbf{log}}: Outputs the current fact base and spatial relations (e.g., as a JSON data, as markdown file with knowledge graph diagrams, or as 3D scene export) for debugging and visualization.
\end{itemize}

Certain steps (\texttt{filter}, \texttt{pick}, \texttt{select}, \texttt{slice}, \texttt{produce}, \texttt{map}, \texttt{adjust}, and \texttt{reload} \emph{change} the number of objects being processed, while others, \texttt{deduce}, \texttt{sort}, \texttt{calc},  and \texttt{log}) simply reorder or annotate them.

The \texttt{sort} operation supports ascending and descending orders (by the \texttt{<} and \texttt{>} comparator) and beside numeric object attributes may be applied to metric values (distance and angle delta) of spatial relations. The relations for sorting are selected between current and backtraced objects. By default, backtracing goes one step back in the inference pipeline; if necessary, the backtracing steps can be set higher. 

\subsection{Syntax Example}
\label{subsec:pipeline-example}

The pipeline syntax uses the pipe character (`\texttt{|}`) to chain operations. Below is a simple example:

\begin{verbatim}
filter(volume > 0.4) 
| pick(left AND above) 
| log()
\end{verbatim}

\begin{itemize}
    \item \texttt{filter(volume > 0.4)} retains only objects with a calculated volume above 0.4.
    \item \texttt{pick(left AND above)} selects objects that are simultaneously \emph{left of} and \emph{above}  the filtered set of spatial objects, based on deduced relations.
    \item \texttt{log()} writes out a summary of these objects for inspection, either as a console output or as a log file.
\end{itemize}

\section{Spatial Inference Examples}
\label{sec:inference-examples}

The following code examples for spatial inference pipelines 
demonstrate the use and feasibility of the Spatial Reasoner.

\subsection{Property Filters}

\noindent{Select a spatial object by its unique identifiier:}

\begin{verbatim}
filter(id == 'id1234')
\end{verbatim}

\noindent{Filter spatial objects by boolean attributes:}

\begin{verbatim}
filter(virtual AND NOT moving)
\end{verbatim}

\noindent{Filter spatial objects by non-spatial attributes:}

\begin{verbatim}
filter(label == 'table' AND 
       confidence.label > 0.7)
\end{verbatim}

\noindent{Filter spatial objects by type attributes:}

\begin{verbatim}
filter((type == 'chair' OR 
        type == 'table'))
\end{verbatim}

\noindent{Filter objects that belong to a type in a class hierarchy:}

\begin{verbatim}
isa(furniture)
\end{verbatim}

Spatial objects have a \texttt{type} attribute which can be embedded into a class hierarchy to reflect the inheritance relations of the entity. Before using the \texttt{isa(type)} operator, a domain-specific taxonomy in OWL/RDF format\footnote{\url{https://www.w3.org/OWL/}} should be imported.
Thanks to the dynamic loading of an application-specific taxonomy, the Spatial Reasoner can be flexibly adapted to the type and label systems used by the underlying computer vision libraries and machine learning models. The \texttt{isa(type)} operator checks for conformance of the type or label with the class name and with optionally specified synonyms along the class hierarchy within the loaded taxonomy.

\subsection{Spatial Queries}

\noindent{Select spatial objects by their spatial attributes:}

\begin{verbatim}
filter(footprint > 0.5 && height > 1.5)
\end{verbatim}

\noindent{Pick spatial objects by their spatial relations:}

\begin{verbatim}
pick(near AND (left OR behind))
\end{verbatim}

\noindent{Get nearest spatial object from observer:}

\begin{verbatim}
filter(id == 'user') 
| pick(disjoint) 
| sort(disjoint.delta <)
| slice(1)
\end{verbatim}

\noindent{Get second left object of type "picture" seen from observer:}

\begin{verbatim}
deduce(topology)
| filter(id == 'user') 
| pick(ahead AND left AND disjoint) 
| filter(type == 'picture') 
| sort(disjoint.delta > -2)
| slice(2)
\end{verbatim}

\noindent{The sort() operation is backtracing by 2 steps to take as input the 'user' object to compare with the relations between 'user' object and filtered objects with type 'picture'.}

\subsection{Object Classifications}

\noindent{Classify spatial objects by their attributes, for example:}

\begin{verbatim}
filter(height < 0.6 && height > 0.25 && 
       width > 1.5 && length > 1.8)
| map(type = 'double bed'; 
      confidence = 0.5)
\end{verbatim}

\noindent{Classify spatial objects by their attributes and their topological arrangement:}

\begin{verbatim}
filter(height > 1.5 && width > 1.0 && 
       depth > 0.4)
| select(backside ? type == 'wall')
| map(type = 'cabinet'; 
      confidence = 0.75)
\end{verbatim}

\noindent{By mapping a \texttt{type} attribute that exists in a loaded taxonomy, the \texttt{isa(type)} operator will consider type inheritance and synonyms.}

\subsection{Production Rules}

\noindent{Create dublicate of a spatial object:}

\begin{verbatim}
filter(id == '1234')
| produce(copy : label = 'copy'; y = 2.0)
\end{verbatim}

\noindent{Aggregate spatial objects by creating a group that covers all children's bbox. For example, create a room object by grouping the existing walls:}

\begin{verbatim}
filter(type == 'wall')
| produce(group : label = 'room')
\end{verbatim}

\noindent{Generate objects at the position where they are connected (at connectivity relations). E.g., create a conceptual spatial object at the corner of each touching wall and placed on the floor:}

\begin{verbatim}
filter(type == 'wall')
| produce(by : label = 'corner'; h = 0.02)
\end{verbatim}
\section{Implementation}
The Spatial Reasoner framework has been realized and validated with test cases as various cross-platform implementations. The following implementations are available as Open Source libraries:

\begin{itemize}
    \item \textbf{SRswift}\footnote{\url{https://github.com/metason/SRswift}} library in Swift for iOS, macOS and visionOS 
    \item \textbf{SRcsharp}\footnote{\url{https://github.com/NicolasLoth/SRcsharp}} library in C\# with bindings for Unity\footnote{\url{https://github.com/NicolasLoth/SRunity}}
    \item \textbf{SRpy}\footnote{\url{https://github.com/metason/SRpy}} library in Python for server-side processing
\end{itemize}

By maintaining the same inference pipeline definitions in all implementations, Spatial Reasoner supports straightforward cross-platform reuse of spatial knowledge across Python, Swift, and C\#. Future JavaScript implementations of Spatial Reasoner are considered to support WebXR applications.

\begin{figure}[htbp]
\centerline{\includegraphics[width=0.4\textwidth]{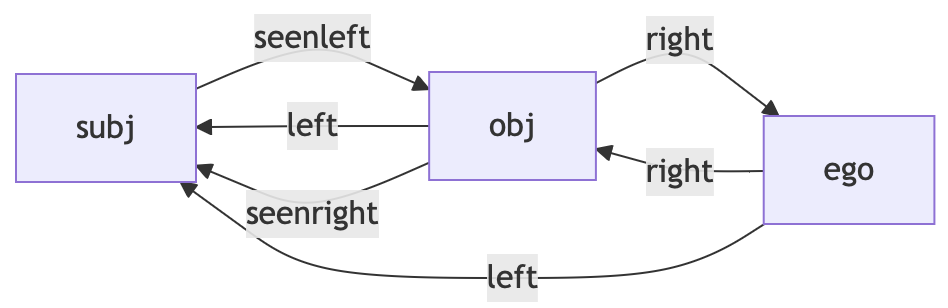}}
\caption{Partial spatial relation graph.}
\label{relationgraph}
\end{figure}

The \texttt{log()} implementations can produce visual graphs of the inferred relationships among spatial objects by creating a Markdown representation using \texttt{mermaid}\footnote{\url{https://github.com/mermaid-js/mermaid}} diagramming syntax. By specifying the relations of interest, e.g., \texttt{log(left right seenleft seenright)}, partial knowledge graphs (such as Figure~\ref{relationgraph}) can be generated for debugging and documentation purposes.  

\begin{figure}[htbp]
\centerline{\includegraphics[width=0.4\textwidth]{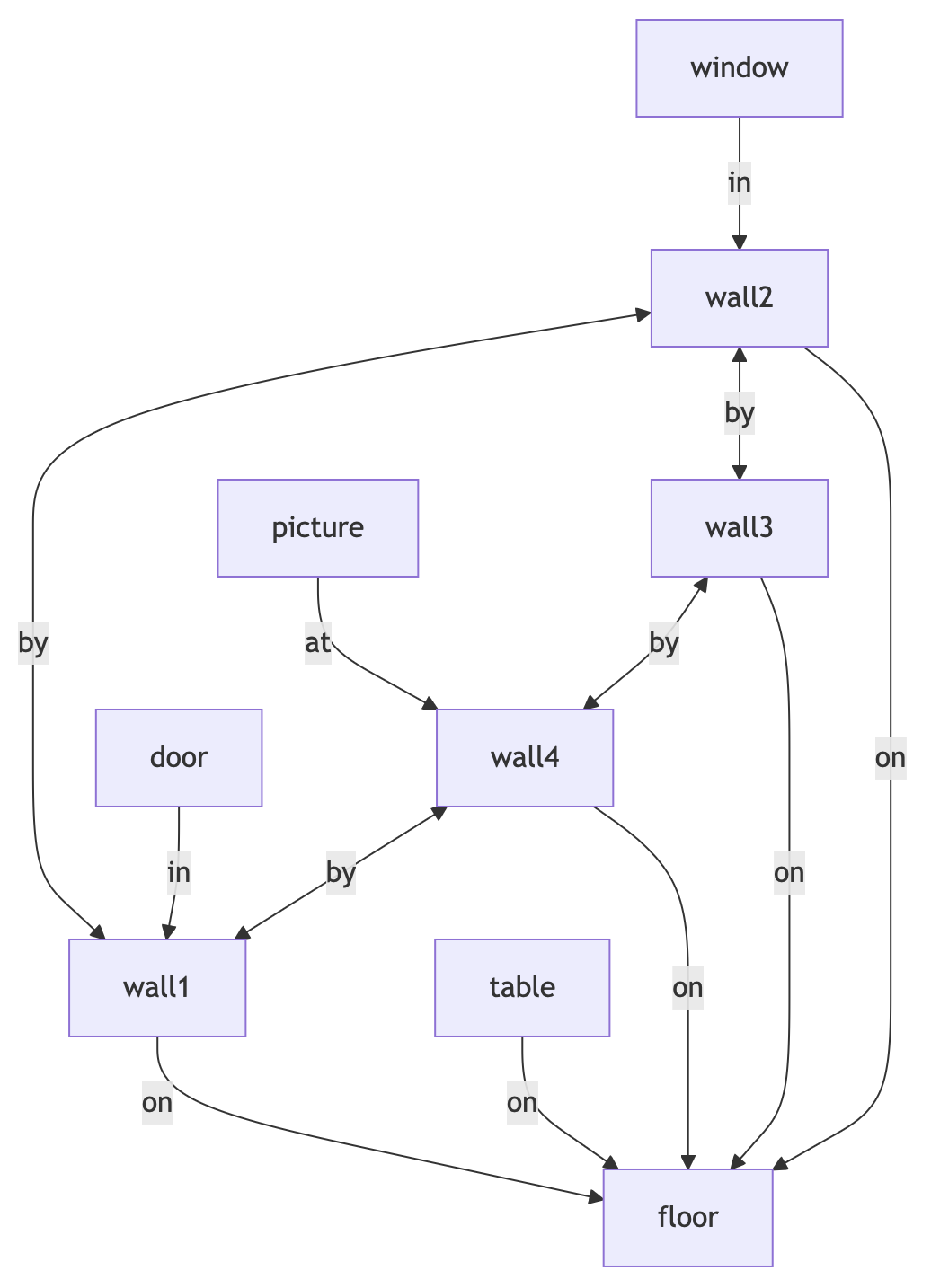}}
\caption{Connectivity graph.}
\label{connectivitygraph}
\end{figure}

Figure~\ref{connectivitygraph} highlights connectivity relations such as \emph{on} or \emph{by}. This allows users to track how objects (e.g., \emph{walls, table, door}) interrelate, and to confirm that spatial relations are deduced as intended.

By calling the \texttt{log(3D)} operation in the spatial inference pipeline, our framework can also generate 3D scenes from objects in the fact base of the Spatial Reasoner. The 3D scenes can be exported for visualization and further analysis (and have been used to create the visuals in Tables~\ref{tab:adjacency} -~\ref{tab:proximity} of the appendix.

\section{Conclusion}
The Spatial Reasoner introduces a robust framework for 3D spatial cognition that seamlessly integrates geometric processing with high-level semantic interpretation. It uses simple, oriented bounding-boxes to represent real and virtual objects, ensuring efficiency and easy integration. At the same time, the surrounding space is divided into discrete sectors to capture nuanced relationships. By employing a rich set of spatial predicates, the framework formalizes relations (e.g., topology, connectivity, and directionality) in a manner that is both metric-driven and semantically informed.

A key strength of the system is its pipeline-based inference architecture, which allows experts to dynamically compose and update rules tailored to specific tasks, such as object classification, scene understanding, user interactions, or game logic in 3D spaces. This design not only streamlines the development of spatial queries, but also supports advanced operations like frequent real-time updates and seamless integration with machine learning systems that generate raw bounding-box detections. By merging lower-level numeric data with higher-level semantics (such as adjacency, connectivity, and visibility) within a single coherent platform, the Spatial Reasoner delivers a versatile solution for a wide range of application domains.

Looking ahead, future work will focus on validating the approach in various XR application scenarios, on creating a spatial ontology editor with run-time testing and monitoring, on extending it to handle spatio-temporal reasoning, and to explore the interplay with natural language processing (NLP) and large language models (LLMs).

\bibliographystyle{IEEEtran}
\bibliography{ref} 
\section{Appendix}

This appendix compiles the key spatial predicates used throughout our framework. Each subsection presents a category of predicates with a table summarizing (i) the predicate name, (ii) the spatial relation it encodes, (iii) a concise specification, and (iv) an illustrative visual.

\begin{table}[!htb]
\centering      
\caption{\textbf{Directionality predicates} indicate left/right or front/back type relationships. They can also define above/over and below/under in world or local object coordinates.}
\label{tab:directionality}
\begin{tabular}{M{1.0cm} L{1.2cm} L{2.6cm} M{2.3cm}}
\toprule
\textbf{Predicate} & \textbf{Relation} & \textbf{Specification} & \textbf{Visual} \\
\midrule
\texttt{left}
& subj is \textbf{left} of obj
& \begin{tblitemize}
    \item center of subject is left
    \item may overlap, no distance condition
    \item valid in OCS
\end{tblitemize}
& \includegraphics[width=2.3cm]{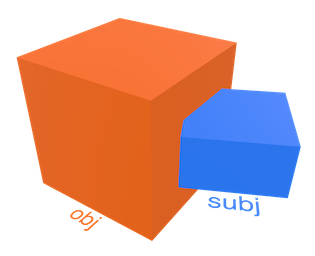} \\
\texttt{right}
& subj is \textbf{right} of obj
& \begin{tblitemize}
    \item center of subject is right
    \item may overlap, no distance condition
    \item valid in OCS
\end{tblitemize}
& \includegraphics[width=2.1cm]{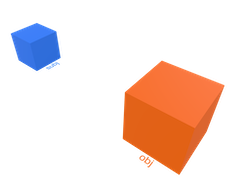} \\
\texttt{ahead}
& subj is \textbf{ahead} of obj
& \begin{tblitemize}
    \item center of subject is ahead
    \item may overlap, no distance condition
    \item valid in OCS
\end{tblitemize}
& \includegraphics[width=2.1cm]{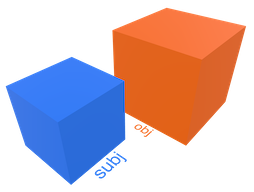} \\
\texttt{behind}
& subj is \textbf{behind} obj
& \begin{tblitemize}
    \item center of subject is behind
    \item may overlap, no distance condition
    \item valid in OCS
\end{tblitemize}
& \includegraphics[width=2.1cm]{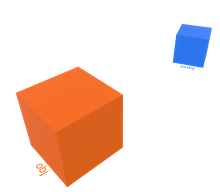} \\
\texttt{above}/ \texttt{over}
& subj is \textbf{above/ over} obj
& \begin{tblitemize}
    \item center is above object
    \item no distance condition
    \item valid in WCS/OCS/ECS
\end{tblitemize}
& \includegraphics[width=1.8cm]{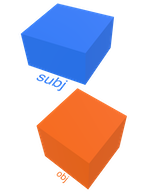} \\
\texttt{below}/ \texttt{under}
& subj is \textbf{below/ under} obj
& \begin{tblitemize}
    \item center is below object
    \item no distance condition
    \item valid in WCS/OCS/ECS
\end{tblitemize}
& \includegraphics[width=1.8cm]{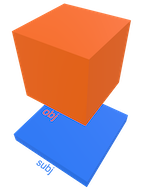} \\
\bottomrule
\end{tabular}
\end{table}
\clearpage
\begin{table*}[p]
\centering 
\begin{minipage}[t]{0.48\textwidth}
\centering
\captionof{table}{\textbf{Adjacency predicates} concern objects that are close but not overlapping, with specification for leftside, rightside, and other relative positions.}
\label{tab:adjacency}
\begin{tabular}{M{1.4cm} L{1.2cm} L{2.2cm} M{2.0cm}}
\toprule
\textbf{Predicate} & \textbf{Relation} & \textbf{Specification} & \textbf{Visual} \\
\midrule
\texttt{leftside} 
& subj is at \textbf{leftside} of obj 
& \begin{tblitemize}
        \item center is in .l sector
        \item is near, not overlapping
        \item valid in OCS
    \end{tblitemize}
& \includegraphics[width=2.0cm]{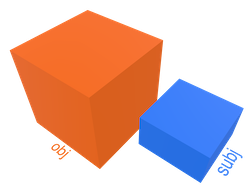} \\
\texttt{rightside} 
& subj is at \textbf{rightside} of obj 
& \begin{tblitemize}
    \item center is in .r sector
    \item is near, not overlapping
    \item valid in OCS
\end{tblitemize} 
& \includegraphics[width=2.0cm]{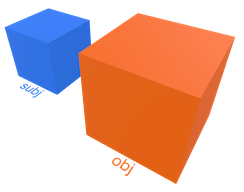} \\
\texttt{frontside} 
& subj is at \textbf{frontside} of obj 
& \begin{tblitemize}
    \item center in .a sector
    \item near, not overlapping
    \item valid in OCS
\end{tblitemize} 
& \includegraphics[width=2.0cm]{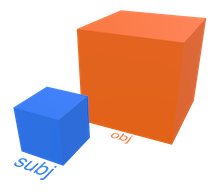} \\
\texttt{backside} 
& subj is at \textbf{backside} of obj 
& \begin{tblitemize}
    \item center in .b sector
    \item near, not overlapping
    \item valid in OCS
\end{tblitemize} 
& \includegraphics[width=2.0cm]{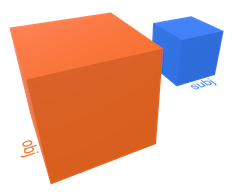} \\
\texttt{beside} 
& subj is \textbf{beside} obj 
& \begin{tblitemize}
    \item near, not above/below
    \item not overlapping
    \item valid in WCS/OCS/ECS
\end{tblitemize} 
& \includegraphics[width=2.0cm]{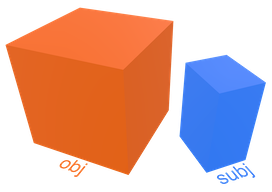} \\
\texttt{upperside} 
& subj is at \textbf{upperside} of obj 
& \begin{tblitemize}
    \item center in .o sector
    \item near, not overlapping
    \item valid in WCS/OCS/ECS
\end{tblitemize} 
& \includegraphics[width=2.0cm]{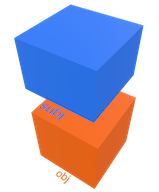} \\
\texttt{lowerside} 
& subj is at \textbf{lowerside} of obj 
& \begin{tblitemize}
    \item center in .u sector
    \item near, not overlapping
    \item valid in WCS/OCS/ECS
\end{tblitemize} 
& \includegraphics[width=2.0cm]{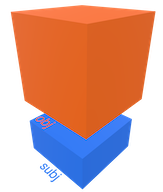} \\
\texttt{ontop} 
& subj is \textbf{ontop} of obj 
& \begin{tblitemize}
    \item center in .o sector
    \item near, not overlapping
    \item min dist $<$ max gap
\end{tblitemize} 
& \includegraphics[width=2.0cm]{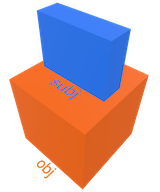} \\
\texttt{beneath} 
& subj is \textbf{beneath} obj 
& \begin{tblitemize}
    \item center in .u sector
    \item near, not overlapping
    \item min dist $<$ max gap
\end{tblitemize} 
& \includegraphics[width=2.0cm]{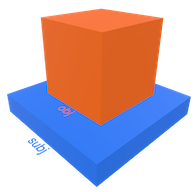} \\
\bottomrule
\end{tabular}
\end{minipage}
\hfill
\begin{minipage}[t]{0.48\textwidth}
\centering
\captionof{table}{\textbf{Assembly predicates} describe overlap, containment, and adjacency in terms of bounding-box relationships of spatial objects.}
\label{tab:assembly}
\begin{tabular}{M{1.6cm} L{1.5cm} L{2.2cm} M{2.0cm}}
\toprule
\textbf{Predicate} & \textbf{Relation} & \textbf{Specification} & \textbf{Visual} \\
\midrule
\texttt{disjoint}
& subj is \textbf{disjoint} from obj
& \begin{tblitemize}
    \item no overlap
    \item valid in WCS/OCS/ECS
\end{tblitemize}
& \includegraphics[width=2.0cm]{images/ahead.png} \\
\texttt{inside}
& subj is \textbf{inside} obj
& \begin{tblitemize}
    \item bounding box inside
    \item valid in WCS/OCS/ECS
\end{tblitemize}
& \includegraphics[width=2.0cm]{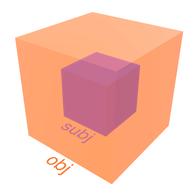} \\
\texttt{containing}
& subj is \textbf{containing} obj
& \begin{tblitemize}
    \item is containing
    \item valid in WCS/OCS/ECS
\end{tblitemize}
& \includegraphics[width=2.0cm]{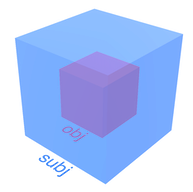} \\
\texttt{overlapping}
& subj is \textbf{overlapping} with obj
& \begin{tblitemize}
    \item partial overlap
    \item not crossing fully
\end{tblitemize}
& \includegraphics[width=2.0cm]{images/left.png} \\
\texttt{crossing}
& subj is \textbf{crossing} obj
& \begin{tblitemize}
    \item is crossing
    \item valid in WCS/OCS/ECS
\end{tblitemize}
& \includegraphics[width=2.0cm]{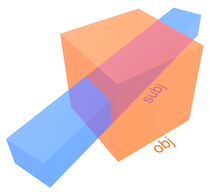} \\
\texttt{touching}
& subj is \textbf{touching} obj
& \begin{tblitemize}
    \item edge/corner contact
    \item min distance $<$ max gap
\end{tblitemize}
& \includegraphics[width=2.0cm]{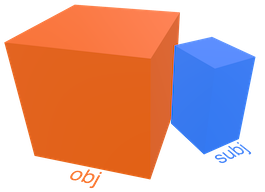} \\
\texttt{meeting}
& subj is \textbf{meeting} obj
& \begin{tblitemize}
    \item face contact
    \item angle diff $<$ max angle
\end{tblitemize}
& \includegraphics[width=2.0cm]{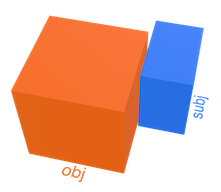} \\
\bottomrule
\end{tabular}%
\vspace{0.3cm}
\captionof{table}{\textbf{Sectoriality predicates} split the space around an object into finer “3D sectors” (e.g., bru standing for behind-right-under) for more granular positional reasoning}
\label{tab:sectoriality}
\begin{tabular}{M{0.8cm} L{1.5cm} L{2.2cm} M{2.0cm}}
\toprule
\textbf{Predicate} & \textbf{Relation} & \textbf{Specification} & \textbf{Visual} \\
\midrule
\texttt{o}
& subj in sector \textbf{o} of obj
& \begin{tblitemize}
    \item center in .o
    \item is over
    \item valid in OCS
\end{tblitemize}
& \includegraphics[width=2.0cm]{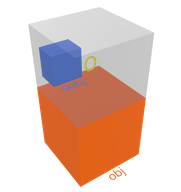} \\
\texttt{br}
& subj in sector \textbf{br} of obj
& \begin{tblitemize}
    \item center in .br
    \item behind right
    \item valid in OCS
\end{tblitemize}
& \includegraphics[width=2.0cm]{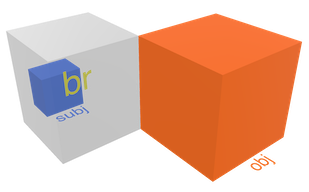} \\
\texttt{bru}
& subj in sector \textbf{bru} of obj
& \begin{tblitemize}
    \item center in .bru
    \item behind right under
    \item valid in OCS
\end{tblitemize}
& \includegraphics[width=2.0cm]{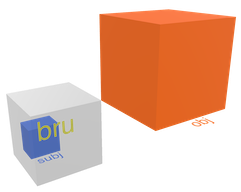} \\
\bottomrule
\end{tabular}%
\end{minipage}
\end{table*}
\clearpage
\begin{table*}[p]
\centering 
\begin{minipage}[t]{0.48\textwidth}
\centering
\captionof{table}{\textbf{Connectivity predicates} describe direct contacts like being “on” or “in,” considering minimal gaps and bounding-box containment.}
\label{tab:connectivity}
\begin{tabular}{M{1.0cm} L{1.0cm} L{2.9cm} M{2.0cm}}
\toprule
\textbf{Predicate} & \textbf{Relation} & \textbf{Specification} & \textbf{Visual} \\
\midrule
\texttt{on}
& subj is \textbf{on} obj
& \begin{tblitemize}
    \item is on top
    \item near, min distance $<$ max gap
    \item valid in WCS/OCS/ECS
\end{tblitemize}
& \includegraphics[width=2.0cm]{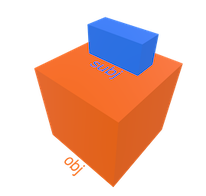} \\
\texttt{at}
& subj is \textbf{at} obj
& \begin{tblitemize}
    \item is beside
    \item is meeting
    \item min distance $<$ max gap
\end{tblitemize}
& \includegraphics[width=2.0cm]{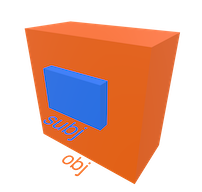} \\
\texttt{by}
& subj is \textbf{by} obj
& \begin{tblitemize}
    \item is touching
    \item min distance $<$ max gap
\end{tblitemize}
& \includegraphics[width=2.0cm]{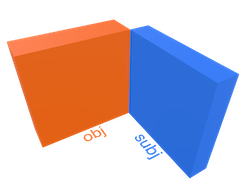} \\
\texttt{in}
& subj is \textbf{in} obj
& \begin{tblitemize}
    \item bounding box inside
    \item valid in WCS/OCS/ECS
\end{tblitemize}
& \includegraphics[width=2.0cm]{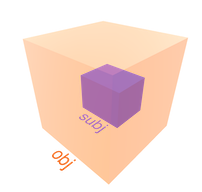} \\
\bottomrule
\end{tabular}%
\vspace{1cm}
\captionof{table}{\textbf{Visibility predicates} focus on how an object is perceived from a specific viewpoint or “ego” coordinate system.}
\label{tab:visibility}
\begin{tabular}{M{1.3cm} L{1.4cm} L{2.0cm} M{2.2cm}}
\toprule
\textbf{Predicate} & \textbf{Relation} & \textbf{Specification} & \textbf{Visual} \\
\midrule
\texttt{infront}
& subj is \textbf{in front} of obj
& \begin{tblitemize}
    \item seen in front
    \item valid in ECS
\end{tblitemize}
& \includegraphics[width=2.2cm]{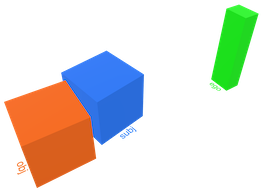} \\
\texttt{atrear}
& subj is \textbf{at rear} of obj
& \begin{tblitemize}
    \item seen behind
    \item valid in ECS
\end{tblitemize}
& \includegraphics[width=2.2cm]{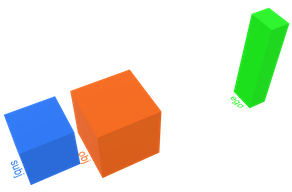} \\
\texttt{seenright}
& subj is \textbf{seen right} of obj
& \begin{tblitemize}
    \item angle from observer
    \item valid in ECS
\end{tblitemize}
& \includegraphics[width=2.0cm]{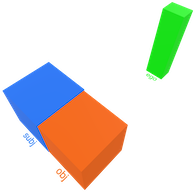} \\
\texttt{seenleft}
& subj is \textbf{seen left} of obj
& \begin{tblitemize}
    \item angle from observer
    \item valid in ECS
\end{tblitemize}
& \includegraphics[width=2.2cm]{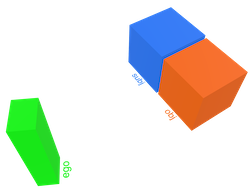} \\
\bottomrule
\end{tabular}%
\end{minipage}%
\hfill
\begin{minipage}[t]{0.48\textwidth}
\centering
 \captionof{table}{\textbf{Proximity predicates} capture whether two objects are near or far from each other, typically comparing center distances against a predefined threshold.}
\label{tab:proximity}
\begin{tabular}{M{1.0cm} L{1.1cm} L{2.9cm} M{2.0cm}}
\toprule
\textbf{Predicate} & \textbf{Relation} & \textbf{Specification} & \textbf{Visual} \\
\midrule
\texttt{near} 
& subj is \textbf{near} obj
& \begin{tblitemize}
    \item center distance $<$ nearby condition
    \item not in \texttt{.i} sector
    \item \texttt{delta} = center distance
\end{tblitemize}
& \includegraphics[width=2.0cm]{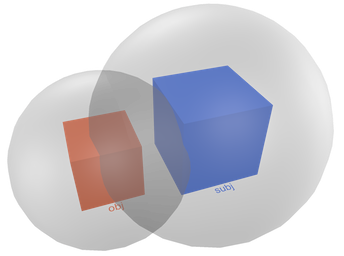} \\
\texttt{far} 
& subj is \textbf{far} from obj
& \begin{tblitemize}
    \item center distance $>$ nearby condition
    \item not near
    \item \texttt{delta} = center distance
\end{tblitemize}
& \includegraphics[width=2.0cm]{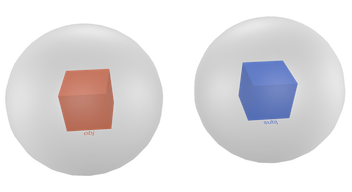} \\
\bottomrule
\end{tabular}%
\vspace{1cm}
\centering
\renewcommand{\arraystretch}{1.5}
\captionof{table}{\textbf{Comparability Predicates} (Compact Version) compare two objects with respect to size, length, volume, and so on.
}
\label{tab:comparability}
\begin{tabular}{M{1.4cm} L{2.9cm} P{3cm}}
\toprule
\textbf{Predicate} & \textbf{Relation} & \textbf{Specification} \\
\midrule
\texttt{shorter}
  & subj is \textbf{shorter} than obj
  & length is shorter; \texttt{delta} = height difference \\[2pt]
\texttt{longer}
  & subj is \textbf{longer} than obj
  & length is larger; \texttt{delta} = length difference \\[2pt]
\texttt{taller}
  & subj is \textbf{taller} than obj
  & height is larger; \texttt{delta} = height difference \\[2pt]
\texttt{thinner}
  & subj is \textbf{thinner} than obj
  & footprint is smaller; \texttt{delta} = footprint difference \\[2pt]
\texttt{wider}
  & subj is \textbf{wider} than obj
  & footprint is larger; \texttt{delta} = footprint difference \\[2pt]
\texttt{smaller}
  & subj is \textbf{smaller} than obj
  & volume is smaller; \texttt{delta} = volume difference \\[2pt]
\texttt{bigger}
  & subj is \textbf{bigger} than obj
  & volume is bigger; \texttt{delta} = volume difference \\[2pt]
\texttt{fitting}
  & subj is \textbf{fitting} into obj
  & is fitting within; \texttt{delta} = volume difference \\[2pt]
\texttt{exceeding}
  & subj is \textbf{exceeding} obj
  & not fitting within; \texttt{delta} = volume difference \\
\bottomrule
\end{tabular}
\end{minipage}
\end{table*}
\clearpage
\begin{table}[!htb]
\centering
\renewcommand{\arraystretch}{1.5}
\caption{\textbf{Similarity Predicates} (Compact Version) check whether objects share dimensions, positions, or surface properties within a chosen tolerance.}
\label{tab:similarity}
\begin{tabular}{M{1.9cm} L{1.9cm} P{3.6cm}}
\toprule
\textbf{Predicate} & \textbf{Relation} & \textbf{Specification} \\
\midrule
\texttt{sameheight}   & subj has \textbf{sameheight} as obj   & height difference $<\!$ max gap \\[2pt]
\texttt{samewidth}    & subj has \textbf{samewidth} as obj    & width difference $<\!$ max gap \\[2pt]
\texttt{samedepth}    & subj has \textbf{samedepth} as obj    & depth difference $<\!$ max gap \\[2pt]
\texttt{samelength}   & subj has \textbf{samelength} as obj   & length difference $<\!$ max gap \\[2pt]
\texttt{sameperimeter}& subj has \textbf{sameperimeter} as obj& perimeter diff $< 4\times$(max gap) \\[2pt]
\texttt{samefront}    & subj has \textbf{samefront} as obj    & front area diff $<\!$ (max gap)$^2$ \\[2pt]
\texttt{sameside}     & subj has \textbf{sameside} as obj     & side area diff $<\!$ (max gap)$^2$ \\[2pt]
\texttt{samefootprint}& subj has \textbf{samefootprint} as obj& base area diff $<\!$ (max gap)$^2$ \\[2pt]
\texttt{samesurface}  & subj has \textbf{samesurface} as obj  & surface diff $< 3\times$(max gap)$^2$ \\[2pt]
\texttt{samevolume}   & subj has \textbf{samevolume} as obj   & volume diff $<\!$ (max gap)$^3$ \\[2pt]
\texttt{samecuboid}   & subj has \textbf{samecuboid} as obj   & same width, height, and depth \\[2pt]
\texttt{congruent}    & subj is \textbf{congruent} with obj   & same dimensions + orientation \\[2pt]
\texttt{sameposition} & subj has \textbf{sameposition} as obj & position diff $<\!$ max gap \\[2pt]
\texttt{samecenter}   & subj has \textbf{samecenter} as obj   & center diff $<\!$ max gap \\[2pt]
\texttt{sameshape}    & subj has \textbf{sameshape} as obj    & same shape label \\
\bottomrule
\end{tabular}
\end{table}

\begin{table*}[!htb]
\centering
\renewcommand{\arraystretch}{2.0}
\caption{\textbf{Available operations} in inference pipeline of Spatial Reasoner}
\label{tab:inference-ops}
\begin{tabular}{p{1.3cm} p{7.3cm} p{7.5cm}}
\hline
\textbf{Op} & \textbf{Syntax} & \textbf{Examples} \\
\hline
\texttt{adjust} & \texttt{adjust(\textit{settings})} & \texttt{adjust(max gap 0.05)} \newline \texttt{adjust(sector fixed 1.2)} \newline \texttt{adjust(nearby dimension 2.0)} \newline \texttt{adjust(long ratio 4.0)} \newline\texttt{adjust(nearby limit 4.0; max gap 0.1)}\\
\texttt{deduce} & \texttt{deduce(\textit{relation-categories})} & \texttt{deduce(topology)} \newline \texttt{deduce(visibility comparability)} \newline \texttt{deduce(topology visibility similarity)} \\
\texttt{filter} & \texttt{filter(\textit{attribute-conditions})} & \texttt{filter(id == 'wall2')} \newline \texttt{filter(width > 0.5 AND height < 2.4)} \newline \texttt{filter(type == 'chair')} \newline \texttt{filter(long)} \newline \texttt{filter(thin AND volume > 0.4)} \\
\texttt{isa} & \texttt{isa(\textit{class-type})} & \texttt{isa('Bed')} \newline \texttt{isa(Furniture)} \newline \texttt{isa(Computer OR Monitor)} \\
\texttt{pick} & \texttt{pick(\textit{relation-conditions})} & \texttt{pick(near)} \newline \texttt{pick(ahead AND smaller)} \newline \texttt{pick(near AND (left OR right))}\\
\texttt{select} & \texttt{select(\textit{relations [? attribute-conditions]})} & \texttt{select(opposite)} \newline \texttt{select(ontop ? label == 'table')} \newline  \texttt{select(on ? type == 'floor')} \newline \texttt{select(ahead AND smaller ? footprint < 0.5)} \\
\texttt{sort} & \texttt{sort(\textit{object-attribute [comparator]})} & \texttt{sort(width)} \newline \texttt{sort(volume <)} \newline \texttt{sort(volume >)} \\
\texttt{sort} & \texttt{sort(\textit{relation-attribute [comparator steps]})} & \texttt{sort(near.delta)} \newline \texttt{sort(frontside.angle <)} \newline \texttt{sort(disjoint.delta > -2)} \\
\texttt{slice} & \texttt{slice(\textit{range})} & \texttt{slice(1)} \newline \texttt{slice(2..3)} \newline \texttt{slice(-1)} \\
\texttt{calc} & \texttt{calc(\textit{variable-assignments})} & \texttt{calc(cnt = count(objects))} \newline \texttt{calc(vol = objects[0].volume)} \newline \texttt{calc(maxvol = max(objects.volume))} \\
\texttt{map} & \texttt{map(\textit{attribute-assignments})} & \texttt{map(shape = 'cylindrical')} \newline \texttt{map(type = 'bed'; shape = 'cubical')} \newline \texttt{map(weight = volume * 140.0)} \\
\texttt{produce} & \texttt{produce(\textit{relation : attribute-assignments})} & \texttt{produce(copy : id = 'copy'; y = 2.0)} \newline \texttt{produce(group : type = 'room')} \newline \texttt{produce(on : label = 'zone')} \newline \texttt{produce(by : label = 'corner'; h = 0.02)} \newline \texttt{produce(al : label = 'ahead-left')} \\
\texttt{backtrace} & \texttt{backtrace(\textit{steps})} & \texttt{backtrace(-2)} \\
\texttt{reload} & \texttt{reload()} & \texttt{reload()} \\
\texttt{halt} & \texttt{halt()} & \texttt{halt()} \\
\texttt{log} & \texttt{log(3D base \textit{relations})} & \texttt{log(3D)} \newline \texttt{log(base)} \newline \texttt{log(on at by in)} \newline \texttt{log(3D near left right)} \\
\hline
\end{tabular}
\end{table*}

\clearpage

\end{document}